\documentclass[letterpaper,12pt]{article} 

\usepackage{amsmath}
\usepackage{amssymb}
\usepackage[comma,longnamesfirst]{natbib}
\usepackage[dvips]{epsfig}
\usepackage{dcolumn}
\usepackage{enumerate}
\usepackage{hhline}
\usepackage{dsfont}
\usepackage{afterpage}
\usepackage{arydshln}
\usepackage{graphicx}
\usepackage{color}
\usepackage[usenames,dvipsnames]{xcolor}
\usepackage{rotating}
\usepackage[breaklinks]{hyperref}
\usepackage{breakurl} 
\usepackage{xr}
\usepackage[percent]{overpic}
\usepackage{subfig}
\usepackage[]{caption}

\usepackage{algorithmicx}
\usepackage[noend]{algpseudocode}
\usepackage{algorithm}
\usepackage{diagbox}
\usepackage{graphicx}
\usepackage{wrapfig}
\usepackage{lscape}
\input epsf
\usepackage{fontenc}
\usepackage{setspace}
\usepackage{bm}
\usepackage{slashbox}
\usepackage{lscape}
\usepackage{breakurl} 
\usepackage{multirow}
\usepackage{eurosym}
\usepackage{titlesec}
\epsfverbosetrue
\def\theequation{\thesection.\arabic{equation}}  
\def\abstract{\if@twocolumn
\section*{Abstract}
\else \normalsize 
\begin{center}
{\bf Summary\vspace{-.5em}\vspace{0pt}} 
\end{center}
\quotation 
\fi}
\def\endabstract{\if@twocolumn\else\endquotation\fi}

\makeatletter
\newcommand{\myappendix}[1]{
	\setcounter{section}{1}
        \renewcommand{\thesection}{A\arabic{section}}}

\def \bvec {\text{\boldmath$b$}}

\def \hvec {\text{\boldmath$h$}}

\def \uvec {\text{\boldmath$u$}}

\def \xvec {\text{\boldmath$x$}}    
\def \yvec {\text{\boldmath$y$}}    \def \mY {\text{\boldmath$Y$}}
\def \zvec {\text{\boldmath$z$}}

\def \betavec         {\text{\boldmath$\beta$}}

\def \varepsilonvec   {\text{\boldmath$\varepsilon$}}
\def \zetavec         {\text{\boldmath$\zeta$}}

\def \varthetavec     {\text{\boldmath$\vartheta$}}
\def \iotavec         {\text{\boldmath$\iota$}}

\def \xivec           {\text{\boldmath$\xi$}}

\def \nullvec {\mathbf{0}}

\usepackage{color}
\usepackage{colordvi}
\newlength{\breite}
\breite\textwidth
\newcounter{aufg}[section]
  {\refstepcounter{aufg}\noindent\textbf{Exercise \arabic{aufg}:}
   \\*[1ex]\noindent}{\vspace{.5cm}}
   
 \newcounter{notes}[section]
  {\refstepcounter{aufg}\noindent\textbf{}
   \\*[1ex]\noindent}{\vspace{.5cm}}
   
\usepackage{amsthm}  

\theoremstyle{definition}

\newtheorem*{beisp*}{Example}
\newtheorem{Proof}{Proof}


\newtheoremstyle{break}
  {}
  {}
  {}
  {}
  {\bfseries}
  {.}
  {\newline}
  {}
  
\theoremstyle{break}

\newcommand{\head}[2]%
 {\hrule \vspace{.15cm} {\sfbold Advanced Statistical Inference, Summer Term 2012, Georg-August-University G\"ottingen}\hfill
{\sfbold Sheet #1}\\
{\sfbold Prof. Dr. Thomas Kneib, Nadja Klein}\hfill {\sfbold #2}

\vspace{.2cm}
\hrule

\vspace{1cm}

}

\newcounter{auf}
{\refstepcounter{auf}
\begin{center}
\fcolorbox[gray]{0}{.95}{
\makebox[\breite]{
\textbf{Exercise \arabic{auf}}
}}\\*[1ex]\noindent
\end{center}
}{\vspace{.5cm}}

\newcounter{loes}[section]
{\stepcounter{loes}
\begin{center}
\fcolorbox[gray]{0}{.95}{
\makebox[\breite]{
\textbf{L"osung \arabic{loes}}
}}\\*[1ex]\noindent
\end{center}
}{}

{\begin{center}
\fcolorbox[gray]{0}{.95}{
\makebox[\breite]{
\textbf{Zu Aufgabe #1}
}}\\*[1ex]\noindent
\end{center}\vspace{1cm}
}{\vspace{1cm}}

\newcounter{ka}
{\refstepcounter{ka}
\begin{center}
\framebox[\textwidth]{
\textbf{Aufgabe \arabic{ka}} \hfill #1 Punkte
}\\*[1ex]\noindent
\end{center}
}{\vspace{1cm}}

\newcounter{lka}
{\refstepcounter{lka}
\begin{center}
\framebox[\textwidth]{
\textbf{L\"osung \arabic{lka}} \hfill #1 Punkte
}\\*[1ex]\noindent
\end{center}
}{\vspace{1cm}}

\usepackage[margin=1in]{geometry}
\titlespacing*\section{0pt}{0pt plus 4pt minus 2pt}{0pt plus 2pt minus 2pt}
\titlespacing*\subsection{0pt}{0pt plus 4pt minus 2pt}{0pt plus 2pt minus 2pt}
\titlespacing*\subsubsection{0pt}{0pt plus 4pt minus 2pt}{0pt plus 2pt minus 2pt}

\newcounter{myremark}

\newcounter{mynotation}

\usepackage{paralist}

\renewenvironment{itemize}[1]{\begin{compactitem}#1}{\end{compactitem}}

\makeatletter
\def\@seccntformat#1{\@ifundefined{#1@cntformat}%
	{\csname the#1\endcsname\quad}  
	{\csname #1@cntformat\endcsname}
}
\let\oldappendix\appendix 
\renewcommand\appendix{%
	\oldappendix
	\newcommand{\section@cntformat}{\appendixname~\thesection\quad}
}
\makeatother

\usepackage{titlesec}

\usepackage{scalerel,stackengine}
\stackMath
\newcommand\reallywidehat[1]{%
\savestack{\tmpbox}{\stretchto{%
  \scaleto{%
    \scalerel*[\widthof{\ensuremath{#1}}]{\kern-.6pt\bigwedge\kern-.6pt}%
    {\rule[-\textheight/2]{1ex}{\textheight}}
  }{\textheight}%
}{0.5ex}}%
\stackon[1pt]{#1}{\tmpbox}%
}

\begin{document}
\setlength{\abovedisplayskip}{0.15cm}
\setlength{\belowdisplayskip}{0.15cm}
\pagestyle{empty}
\begin{titlepage}

\title{Deep Distributional Time Series Models and the Probabilistic Forecasting of Intraday Electricity Prices}

\author{Nadja Klein, Michael Stanley Smith, David J. Nott}
\date{}
\maketitle
\noindent
{\small Nadja Klein is Assistant Professor of Applied Statistics and Emmy Noether Research Group Leader in Statistics and Data Science at Humboldt-Universit\"at zu Berlin; Michael Stanley Smith is Professor of Management (Econometrics) at Melbourne Business School, University of Melbourne; David J. Nott is Associate Professor of Statistics and Applied Probability at National University of Singapore. Correspondence should be directed to~Prof.~Dr.~Nadja Klein at Humboldt Universit\"at zu Berlin,
Unter den Linden 6, 10099 Berlin. Email: nadja.klein@hu-berlin.de.

\noindent \textbf{Acknowledgments:} Nadja Klein was supported by the Deutsche Forschungsgemeinschaft (DFG, German research foundation) through the Emmy Noether grant KL 3037/1-1.  David Nott is affiliated with the Operations Research and Analytics Research cluster at the National University of Singapore.
The authors thank Sonnia Fuenteseca for research assistance in 
constructing the demand forecast
data used in Section~6, and Patrick McDermott and Chris Wikle for sharing 
their code.}\\

\newpage
\begin{center}
\mbox{}\vspace{2cm}\\
{\LARGE \title{Deep Distributional Time Series Models and the Probabilistic Forecasting of Intraday Electricity Prices}
}\\
\vspace{1cm}
{\Large Abstract}
\end{center}
\vspace{-1pt}
\onehalfspacing
\noindent
Recurrent neural networks (RNNs) with rich feature vectors of past 
values
can provide accurate point forecasts for series that
exhibit complex serial dependence. 
We propose two approaches to constructing
deep time series probabilistic models
based on a variant of RNN called an echo
state network (ESN).
The first is where the output layer of the ESN has 
stochastic disturbances and
a shrinkage prior for additional regularization. The second approach employs
the implicit copula of an ESN with Gaussian disturbances,
which is a deep copula process on the feature space. 
Combining this copula with a non-parametrically estimated
marginal distribution produces 
a deep distributional time series model. 
The resulting probabilistic
forecasts are deep functions of the feature vector and also marginally calibrated. 
In both approaches,
Bayesian Markov chain Monte Carlo methods
are used to estimate the models and compute forecasts. The 
proposed models are suitable for the complex
task of forecasting intraday electricity prices. 
Using data from the Australian
National Electricity Market, we show that our deep time series models
provide accurate short term probabilistic price forecasts, with the copula model
dominating. Moreover, 
the models provide a flexible framework for incorporating probabilistic forecasts of electricity demand as
additional features, which
increases upper tail forecast accuracy from the copula model significantly.

\vspace{20pt}
 
\noindent
{\bf Keywords}: Copula, density forecasts,
echo state network, electricity price forecasting,
 marginal calibration, Markov chain Monte Carlo, recurrent neural network. 
\end{titlepage}

\newpage
\pagestyle{plain}
\setcounter{equation}{0}
\renewcommand{\theequation}{\arabic{equation}}

\section{Introduction}\label{sec:intro}
Deep models with rich feature vectors are often very effective in problems which require accurate forecasts~\citep{goodfellow+bc16}. These include financial applications,
such as predicting equity risk premiums and returns~\citep{feng2018,gu2020,gu2020autoencoder}
and bond returns~\citep{bianchi2020}. 
Another financial application where deep models have high potential is 
the forecasting of intraday electricity prices.
Electricity prices exhibit a strong and complex nonlinear serial
dependence, quite unlike security prices, 
and accounting for this is key to obtaining accurate
forecasts~\citep{nowotarski2018,manner2019}. Shallow neural networks (NNs)~\citep{amjady2006,mandal2007}, and more recently deep neural networks (DNNs)~\citep{lago2018,ugurlu2018}, have been shown to capture these nonlinearities well
and produce accurate point forecasts. However, it is the accurate
forecasting of the entire distribution of prices---variously called probabilistic, density 
or distributional forecasting---that is important for both market operators and participants.
Yet, to date, probabilistic forecasts of electricity prices using DNNs
are rare. In this paper, we propose a number of time series probabilistic forecasting models that exploit and extend state-of-the-art deep models,
and apply them to data from the Australian market.

Day-ahead wholesale electricity markets operate throughout the 
world, including in the U.S. and Europe. 
In these markets, generators and distributors 
bid for sale and purchase of electricity at an intraday resolution in an auction
one day prior to transmission. The auction clearing price is widely 
called the
electricity spot price; see~\cite{kirschen2018} for an introduction to such markets.
Accurate price forecasts at an intraday resolution, one or more days ahead, are 
central to both the efficient
operation of the market and profitability of participants. 
Particularly
important are probabilistic forecasts of price, not just the mean, variance or other moments. This is because overall profitability 
of market participants is strongly affected by prices in the tails, which are very heavy in most wholesale markets.


Recurrent neural networks (RNNs) are DNNs tailored to capture temporal behavior, and are suitable for forecasting  
nonlinear time series~\citep[Ch.10]{goodfellow+bc16}. However, RNNs 
typically have a very large number of hidden weights and are difficult 
to train and tune. Therefore,
we use a variant of RNNs called echo state networks (ESNs)~\citep{Jae2007,LukJae2009},
that are flexible and employ a form of regularization that makes them scalable to long series and computationally stable. 
We
build statistical time series models based on ESNs using two approaches. The
first extends that of   
\cite{chatzis2011echo}
 and \cite{mcdermott+w17,mcdermott2019deep}, who use ESNs within statistical models, with the 
output layer coefficients of the hidden state vector estimated using Bayesian methods. In our work we include a shrinkage prior for
these coefficients to provide additional regularization, 
along with
three different additive error distributions for the output layer: Gaussian, skew-normal and skew-$t$. 

Our second 
approach is the main methodological contribution of the paper. It uses the implicit
copula of the time series vector from a Gaussian probabilistic ESN
of the type described above.
By an ``implicit copula'' we mean the copula that is implicit in a
multivariate distribution and that is obtained by inverting the usual expression of Sklar's theorem
as in~\citet[Sec.~3.1]{nelsen06}. This implicit copula is both a deep function of the 
feature vector, and also a ``copula process''
 with the same dimension as the time series vector.
We combine our proposed copula
with a non-parametrically estimated marginal distribution for electricity prices, 
producing a time series model that captures the 
complex serial dependence in the series.
An accurate estimate of the marginal distribution 
ensures ``marginal calibration'', which is where the long
run average of the predictive distributions of the time series variable matches its
observed margin~\citep{gneiting+br07,gneiting2014}. Importantly,
the entire predictive distribution from the copula model is a deep
function of the feature vector. 
We note that our copula
model extends the deep distributional regression 
methodology of~\cite{KleNotSmi2019} to deep distributional time series and ESNs.

In both our deep time series models
the feature vector includes a rich array of past series
values and possibly other variables. 
To regularize these, ESNs use sparse and
randomly assigned fixed weights for the hidden layers of the DNN. For
each of $K$ random configurations of weights, 
Markov chain Monte Carlo (MCMC) is used to estimate the statistical model  
and compute Bayesian predictive distributions. The probabilistic forecasts
are then ensembles of these predictive distributions over the $K$ configurations of weights. 
We show in our empirical work that this ensemble provides for
accurate uncertainty quantification.

We use our deep time series models to forecast intraday electricity prices
in the Australian National Electricity Market (NEM). 
The NEM 
 is one of the earliest established wholesale markets (in 1998),
 comprises approximately 1\% of all Australian economic output, 
 provides publicly available data,
and has a design that is typical of other day-ahead markets. It has five regional price series for which forecasting has been much studied; see~\cite{ignatieva2016}, \cite{smith2018}, \cite{manner2019} and~\cite{han2020} for overviews. 
Using a feature vector with lagged
prices from all five regions, we compute predictions at the half-hourly 
resolution for a 24 hour horizon over 
an eight month validation period during 2019. 
\cite{serinaldi2011,gianfreda2018} and~\cite{narajewski2020}
show the `generalized additive models for location, scale and shape' (GAMLSS)
methodology of~\cite{rigby+s05}
applied to time series allows for the accurate modeling and forecasting of electricity prices,
and we employ this as a benchmark. 
Using contemporary metrics, the probabilistic
ESN with skew-$t$ errors produces more accurate point and probabilistic
forecasts than ESNs with either Gaussian or skew-normal errors,
and also the GAMLSS benchmark. However, 
the deep copula model produces substantially more accurate probabilistic forecasts, including
in both tails, and its forecasts have superior coverage.
Thus, marginal calibration also improves calibration of
the (conditional) predictive distributions.

Participants in the NEM are provided with  probabilistic forecasts of electricity demand by the system operator.
Recent studies~\citep{ziel2016,shah2020} suggest 
that using accurate demand forecasts may further improve time 
series forecasts of the distribution of price.
An advantage of deep models is that additional predictors
are easily included in a flexible fashion 
as extra elements in the feature vector.
We do so here using three quantiles of the 24 hour ahead demand forecasts, and find 
this increases forecast accuracy of the 
upper tail of the price distribution, but only when using the deep copula model.
Accurate forecasting of the upper tail of electricity prices impacts the
profitability of participants substantially~\citep{christensen2012}.

The paper is organized as follows. Sec.~\ref{sec:markets}
provides an overview of electricity markets and 
price forecasting, with a
focus on the NEM. Sec.~\ref{sec:DNN} outlines ESNs, their probabilistic extension
using additive disturbances, and
Bayesian methods for their estimation and prediction. 
Sec.~\ref{sec:copulamod} outlines
the implicit copula process and the proposed deep distributional time series model. 
Sec.~\ref{sec:empirical} compares
the deep time series and benchmark model forecasts, Sec.~\ref{sec:demand} considers the inclusion of demand forecast data, and 
Sec.~\ref{sec:conc} concludes. The Web Appendix (``WA'' hereafter) provides key algorithms, computational
details and additional empirical results.

\section{Electricity Markets, Price Forecasting and Data}\label{sec:markets}
\subsection{Wholesale markets}
Wholesale electricity markets include the European Power Exchange,
mutiple regional markets in
the U.S. (such as the PJM interconnection and the Southwest Power Pool), and national markets in many countries including Australia,
Chile and Turkey. While the designs of these markets differ, they 
are largely ``day-ahead'' markets where
generators and distributors
place bids for the sale and purchase of electricity at an intraday resolution up to one day prior to
 transmission (or ``dispatch'').
The market is cleared at a wholesale spot price
that reflects the marginal cost of supply at each intraday
period. Prices also vary at different
geographic reference nodes, creating multiple related price series.
Markets are overseen by  system operators,
which match generation with short-term demand forecasts, 
impose constraints to ensure system stability (i.e. avoid load-shedding or blackouts), 
and enforce any price caps; see~\cite{kirschen2018}
for an overview of wholesale markets.

Central to the operation of day-ahead markets is
the intraday electricity spot price. For market participants 
accurate short-term forecasts of the price 
are key to profitability. Because
electricity is a flow commodity with a high cost of storage, 
arbitrage opportunities are limited. 
This fact, along
with the complexities of transmission and that short-run 
demand is 
inelastic with respect to price,
means that prices exhibit unique stylized characteristics;
see~\cite{knittel2005,karakatsani2008,panagiotelis2008} and~\cite{weron2014} for summaries of these.
From a time series perspective, this includes
strong and  complex nonlinear serial dependence, while from a distributional perspective prices
have very heavy tails,  skew and often multiple modes that correspond to different
regimes~\citep{janczura2010}
and economic equilibria~\citep{smith2018} . 

\subsection{Electricity price forecasting}
Many methods
have been used for short-term
forecasting of electricity spot prices; see 
\cite{weron2014} and~\cite{nowotarski2018} for recent
overviews of point
and probabilistic forecasting methods, respectively.
In the machine learning literature, while
shallow neural networks (NNs) have long been popular
for forecasting electricity prices, DNNs have the potential to produce more accurate
forecasts. For example,~\cite{lago2018} and~\cite{ugurlu2018} both
found that RNN models provide more accurate point forecasts
than a range of benchmark models. However, most previous usages of shallow and deep NNs have 
focused on point forecasts of prices, and only in a few cases also 
quantify predictive uncertainty using bootstrap or other Monte Carlo 
methods~\citep{rafiei2016}.

In contrast, a number of other methods have been used to construct probabilistic 
forecasts~\citep{misiorek2006,panagiotelis2008,huurman2012,bunn2016}. One particularly
promising avenue is to extend
distributional regression methods 
to time series forecasting.  
For example, \cite{gianfreda2018} and~\cite{narajewski2020} do so
for German electricity prices using exogenous covariates, and~\cite{serinaldi2011} does so for Californian and Italian electricity prices using historical prices and other variables. These
papers report more accurate probabilistic forecasts.
In Sec.~\ref{sec:copulamod}
we develop a copula-based model that exploits the
accuracy exhibited by deep models within a distributional time 
series forecasting setting, thereby combining the advantages of both approaches.

Copulas have become increasingly popular in time series models
of electricity prices because they can capture
complex dependence, while also allowing
for highly flexible margins. 
\cite{smith2012},
\cite{ignatieva2016}, \cite{manner2016}, \cite{pircalabu2017}
and~\cite{manner2019} use low-dimensional copulas to capture cross-sectional
dependence between regional prices, price spikes and other
energy series in interconnected power systems,
while~\cite{smith2018} 
use high-dimensional copulas
to capture both serial and cross-sectional 
dependence jointly for multiple regional prices. 
However, these studies all employ copulas
that are very different
to the copula process proposed here. 

\subsection{Australian electricity prices}\label{sec:ausdata}
Electricity generation in the NEM is an important component of economic activity,
with 19.4bn Australian dollars of turnover during the 2018-2019
financial year~\citep{AERstate}.  
Operations in the NEM are managed
by the Australian Electricity Market Operator (AEMO), and since
April 2006 it has had
five regions which correspond to the power systems in the 
states of New South Wales (NSW), 
Queensland (QLD), Victoria (VIC),
South Australia (SA) and Tasmania (TAS). Separate prices are set 
at a central location (or ``node'') in each region, although they are dependent
because the state-based power systems are interconnected
by high voltage direct current lines. 

Participating utilities
place bids for the purchase (by distributors) and sale (by generators)
of electricity at five minute intervals up to one day prior to dispatch. 
The trading price is the average clearing price for this auction 
over six consecutive
five minute periods, so that it is observed at a half-hourly resolution.
Re-bidding of prices is allowed before dispatch, 
although not the amount of energy.
Price forecasting in the NEM has
been studied extensively, with contributions by~\cite{higgs2009}, \cite{panagiotelis2008}, 
\cite{nowotarski2013}, 
\cite{ignatieva2016}, \cite{rafiei2016}, \cite{smith2018},
\cite{apergis2019} and~\cite{manner2019} among others. 
Intraday forecasts
over a horizon of 24 hours are used by market participants to develop effective
strategies for bidding, re-bidding and managing risk.
In our study, we employ the half-hourly trading prices 
(measured in Australian dollars per MW/h) observed during 2019 in the five regions 
of the NEM.
Prices can be negative for short periods (when it is more cost-effective to 
sell into the market at a loss, rather than ramp down generation temporarily)
although the floor price is -\$1,000. There is also a maximum 
price that 
is adjusted annually on 1 July, which was \$14,500 prior
to 1 July 2019 and \$14,700 afterwards.

\begin{figure}[ht]
	\caption{Boxplots of the logarithm of electricity prices in the NSW region by hour.}
	\begin{center}
\centering\includegraphics[width=0.5\textwidth,angle=0]{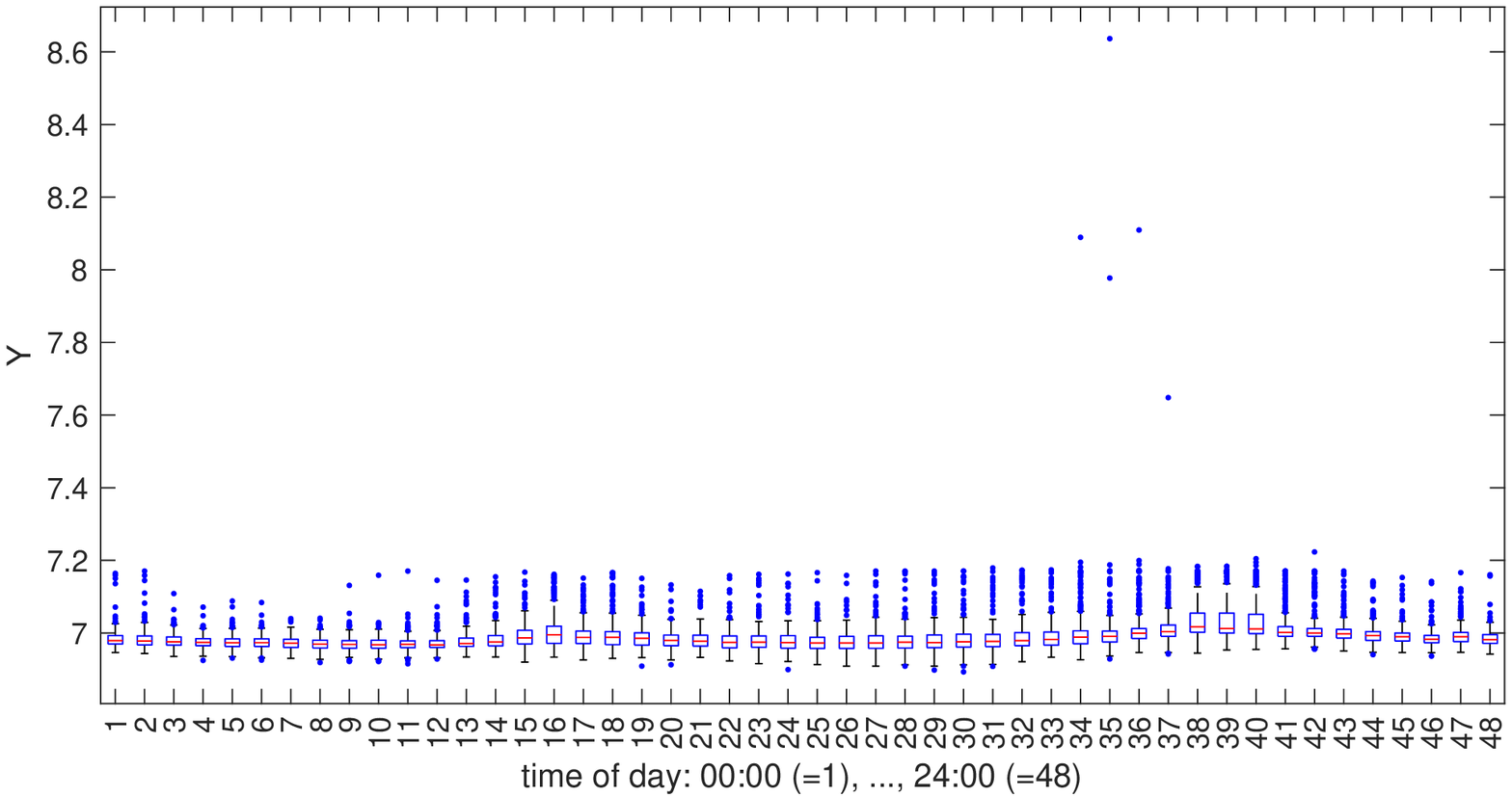}
\end{center}	
Boxplots are of $Y=\log(\mbox{PRICE}+1001)$ for each hour of the day during the period 1 January 2019 to 31 December 2019.
\label{fig:boxplots}
\end{figure}

Fig.~A in the WA~plots prices in NSW during four weeks corresponding
to the four seasons, illustrating the strong 
heterogeneity based on the time of day, day of the week and season.
Table~B in the WA provides a
summary of the prices in each region.
These have extreme positive skew, so that we follow most previous
studies and work with the logarithm of price $Y=\log(\mbox{Price}+1001)$. The addition of 1001 accounts for
the minimum price, at which $Y=0$. 
It is straightforward to construct density forecasts of the 
nominal price from those of $Y$ using the 
Jacobian of the transformation. 
Fig.~B in the WA plots histograms of the five price series,
along with (bounded) kernel density estimates (KDEs), 
showing that even the distribution of $Y$ is both asymmetric and heavy-tailed for each series.
Figure~\ref{fig:boxplots} gives boxplots of $Y$ for NSW (which is the 
region with the most energy demand), broken down by hour of the day. It shows that
the entire distribution varies substantially over the day. Equivalent plots for the day of the week, and month of the year, (see the WA) also reveal similar heterogeneity in the distribution of prices.
\section{Deep Time Series}\label{sec:DNN}
RNNs represent complex
temporal relationships between variables
by allowing cycles and sequences in their hidden layers.
However, RNNs are typically computationally expensive to estimate and can be 
numerically unstable for long series~\citep{PasMikBen2013}, both of which 
are issues here.
One way to
address this is to employ a variant called an Echo State Network (ESN) 
that decreases the number of weights
that need to be trained~\citep{Jae2007}. The key idea of ESNs is to 
only adapt the output layer in training, while keeping the
weights of recurrent and input connections fixed yet randomly assigned. 
Such an approach is called a `reservoir computing' method \citep{LukJae2009}, because
it establishes a multiple linkage hidden reservoir which can be of much higher dimension
than the input, providing a lot of flexibility at a lower computational cost. 
However, classical ESNs and many of its extensions rarely consider uncertainty quantification, and in this
section we outline probabilistic ESNs that do so.
The approach is extended
in Sec.~\ref{sec:copulamod} to allow for marginal calibration 
of the data distribution using a deep copula construction.

\subsection{Gaussian probabilistic ESN}\label{subsec:ESN}
\subsubsection{Specification}
We first specify the ensemble ESN proposed for spatio-temporal data by~\cite{mcdermott+w17}. 
For series with highly nonlinear dependence, these authors demonstrate
that allowing the response equation 
to depend on the hidden states quadratically in an ESN can increase predictive accuracy, which we also find for the electricity price series examined here.
Let $\{Y_t\}$ be a stochastic process,
then the ESN with Gaussian disturbances $\varepsilon_t\sim\mathcal{N}(0,\sigma^2)$ takes the following form for $t\geq 1$: 
\begin{equation}\begin{aligned}\label{eq:ESN}
&\mbox{Response Equation:\quad} &Y_t &= \beta_0+\hvec_t'\betavec_1 + (\hvec_t')^{\circ 2}\betavec_2+\varepsilon_t\,,\\
&\mbox{Hidden State Equation:\quad } &\hvec_t& =(1-\kappa)\hvec_{t-1}+\kappa\tilde \hvec_{t}\\
& &\tilde\hvec_t&= g_h\left(\frac{\delta}{\lambda_V}V\hvec_{t-1}+U \xvec_t\right)\,.
\end{aligned}\end{equation}
Here, $\xvec_t$ is an $n_x$-dimensional feature vector that includes unity for an intercept term, $\hvec_t$ is
a $n_h$-dimensional hidden state vector, `$\circ 2$' denotes the
 element-wise square of a matrix, $V,U$ are $(n_h\times n_h)$ and $(n_h\times n_x)$ matrices of hidden layer weights,
and $g_h$ is an activation function, set here to $g_h(x)=\tanh(x)$. 
The parameter $0<\kappa\leq 1$ and is known as the ``leaking rate''.
Similar to~\cite{mcdermott+w17} we set $\kappa=1$ after having checked that predictive performance is not
improved by setting $\kappa<1$, so that $\tilde \hvec_t=\hvec_t$. The
feature vector $\xvec_t$ contains past values of both the response and other series, which
we specify in Sec.~\ref{sec:empirical}.
The response $Y_t$ is 
a nonlinear
function not only of 
$\xvec_t$, but also of all previous values 
$\xvec_1,\ldots,\xvec_{t-1}$, so the stochastic process is not Markov.


The constant $\lambda_V$ is the largest
eigenvalue of $V$, and
$0<\delta<1$ a scaling parameter, so that $(\delta/\lambda_V)V$ has spectral radius
less than one. A spectral radius greater than one can result in unstable behavior in the latent states~\citep{LukJae2009}. 
We follow~\cite{mcdermott+w17} and set $\delta=0.35$ and $n_h=120$, although we found the
forecasting results to be insensitive to variations in these settings.

The elements of the matrices $V=\{v_{il}\},U=\{u_{ij}\}$ are assumed to be random, and distributed
independently from mixtures of a uniform distribution 
and a point mass at zero. If $\mathcal{U}(a,b)$ denotes a uniform
distribution over domain $(a,b)$, $\mathcal{B}(\pi)$ denotes a beta distribution with mean $\pi$, and
$\delta_0$ is the Dirac function at zero, then the elements
\begin{eqnarray}
v_{il}&=&\gamma_{il}^v\mathcal{U}(-a_v,a_v)+(1-\gamma_{il}^v)\delta_0\,,\;\;\gamma_{il}^v\sim\mathcal{B}(\pi_v)\,, 
\nonumber \\ 
u_{ij}&=&\gamma_{ij}^u\;\mathcal{U}(-a_u,a_u)+(1-\gamma_{ij}^u)\delta_0\,,\;\; \gamma_{ij}^u\sim\mathcal{B}(\pi_u)\,. \label{eq:dsnUV}
\end{eqnarray} 
We follow~\cite{mcdermott+w17} and set $a_v=a_u=\pi_v=\pi_u=0.1$ after having checked the predictive performance under several alternative settings.

Consider $T$ time series observations $\mY=(Y_1,\ldots,Y_T)'$ of the stochastic process with corresponding $(T\times n_x)$ 
matrix of feature values $X=[\xvec_1|\cdots|\xvec_T]'$. Denote  $\xivec=\lbrace V,U,\kappa,\delta\rbrace$,  $H_{\xi}(X)=[\hvec_1|\cdots|\hvec_T]'$ as the $(T\times n_h)$ matrix of hidden
state values, $B_{\xi}(X)=[\iotavec,H_{\xi}(X)|H_{\xi}(X)^{\circ 2}]$, $\iotavec$ as a vector ones, and $\betavec=(\beta_0,\betavec_1',\betavec_2')'$.
Then
\eqref{eq:ESN} can be written
as the linear model
\begin{equation}\label{eq:ESNmat}
\bm{Y}= B_{\xi}(X)\betavec+\varepsilonvec\,,\;\;\mbox{ } \varepsilonvec=(\varepsilon_1,\ldots,\varepsilon_T)'\sim\mathcal{N}(\nullvec,\sigma^2 I).
\end{equation}
Given $\xivec$, $X$ and $\hvec_0=\bm{0}$, 
the hidden state matrix $H_{\xi}(X)$ is known without error
as the hidden
state vectors can be computed recursively. 
Only $\betavec$ and $\sigma^2$ (which we refer to as model parameters) require estimation, for which we use
their Bayesian posterior distribution.
Differing from~\cite{mcdermott+w17}, we
regularize $\betavec$ by adopting the shrinkage prior
\begin{equation}\label{eq:prior1}
\betavec|\tau^2 \sim\mathcal{N}(\nullvec, P(\tau^2)^{-1})\,,\;
\sigma^2\sim\mathcal{IG}(a,b)\,,
\end{equation}
where $\mathcal{IG}$ denotes an Inverse Gamma distribution.
We found a ridge prior with $P(\tau^2)=\tau^2 I$, 
and hyper-prior $\tau^2\sim\mathcal{IG}(\tilde a,\tilde b)$ with $a=b=\tilde a=\tilde b=0.001$
to work well. The posterior of the
model parameters $\varthetavec=(\betavec, \sigma^2,\tau^2)$ of this regularized linear model 
can be
computed using the standard MCMC sampler at Algorithm~1
in Part~A of the WA.

\subsubsection{Probabilistic forecasts}\label{sec:gaussfore}
Most implementations of ESNs only draw a single set of weights 
from~\eqref{eq:dsnUV}. Here, we follow
\cite{mcdermott+w17,mcdermott2019deep} and simulate $K=100$ matrices $\{V^{k},U^{k}\,; k=1,\ldots,K\}$ from~\eqref{eq:dsnUV}.  
An ensemble is then used to integrate over $U,V$ when constructing the 
probabilistic forecasts. 
If $\xivec^{k}=\lbrace V^{k},U^{k},\kappa,\delta\rbrace$, then 
the density forecast of $Y_{T+h}$ at time $T$ for each point $h=1,\ldots,h_1$ in 
the forecast horizon is
the ensemble 
\begin{equation}\label{eq:ensemble}
f_{T+h|T}(y_{T+h})\equiv\frac{1}{K}\sum_{k=1}^K p^k(y_{T+h}|X,\yvec)\,,
\end{equation}
where the subscript notation indicates $f_{T+h|T}$ is conditional on the filtration 
at time $T$.

The density $p^k$ in~\eqref{eq:ensemble} is the Bayesian posterior predictive density computed 
for configuration $\xivec^k$ as follows.
Let $X_{(t)}\equiv[\xvec_1|\cdots|\xvec_t]'$,
then
\begin{eqnarray}
p^k(y_{T+h}|X,\yvec) &= &\int\int p(y_{T+h}|X_{(T+h)},\varthetavec) 
 p(\xvec_{T+2},\ldots,\xvec_{T+h}|\varthetavec,X,\yvec) \nonumber \\
 & &\;\;\;\;\;\;\;\;p(\varthetavec|X,\yvec)\mathrm{d}\xvec_{T+2}\cdots \mathrm{d}\xvec_{T+h}\,\mathrm{d}\varthetavec\,,\label{eq:pdenscop}
\end{eqnarray}
which is an integral over any unobserved feature values and the 
model parameter posterior.\footnote{It is implicit that all densities in the integrand of~\eqref{eq:pdenscop} are conditional on weight configuration $\xivec^k$.}
From~\eqref{eq:ESN}, the first term in the integrand is the density of a $\mathcal{N}(\hvec_{T+h}'\betavec_1+(\hvec_{T+h}')^{\circ 2}\betavec_2,\sigma^2)$ distribution, where 
$\hvec_{T+h}$ is computed through the recursion of the hidden state 
equation (therefore $\hvec_{T+h}$ is a deep function of 
	$\xvec_1,\ldots,\xvec_{T+h}$). 
The outer integral in $\varthetavec$
can be evaluated by averaging
over draws from the posterior $p(\varthetavec|X,\yvec)$ obtained from running the MCMC sampler.
However, plugging in the posterior mean $E(\varthetavec|X,\yvec)$ computed from
the Monte Carlo sample is much faster, yet can be almost as accurate, so we follow this approach.\footnote{Implementation
	requires running the MCMC sampler and computing the posterior mean a total of
	$K$ times, once for each hidden weight configuration 
	$\xivec^1,\ldots,\xivec^K$.}

In our empirical work $\xvec_t$ contains past values of both the focal price
 and the other four regional prices, so that $\xvec_1,\ldots,\xvec_{T+1}$ are observed at or before time $T$. However, some elements of $\xvec_{T+2},\ldots,\xvec_{T+h}$
are unobserved, and the integrals over these feature vectors in~\eqref{eq:pdenscop} 
are with respect to their unobserved elements only. The integrals are 
computed in a Monte Carlo manner by simulating all five series
values sequentially from their predictive distributions.
Algorithm~2 in Part~B of the WA
simulates 
from the ensemble density~\eqref{eq:ensemble} 
for all five regional price series and over each time point in the forecast
horizon.

\subsection{Skew $t$ probabilistic ESN}\label{subsec:skewt:ESN}

Adopting
Gaussian disturbances to the output layer at~\eqref{eq:ESN} 
is inconsistent with the strong
asymmetry in the empirical distribution of (the logarithm of) prices.
Thus, we also consider the skew-$t$ distribution of~\cite{Azz2003} with location zero for the disturbances, with density
$p(\varepsilon_t;\omega^2,\alpha,\nu) = \frac{2}{\omega}t_\nu \left(\frac{\varepsilon_t}{\omega}\right) T_{\nu+1}\left(
\frac{\alpha\varepsilon_t}{\omega} \sqrt{\frac{\nu+1}{\nu+x_{\varepsilon_t}^2}}\right)$, 
where $t_\nu$ and $T_{\nu}$ are the student $t$ density and distribution
functions.
The parameters $\omega^2,\alpha,\nu$ are  scale, skew and degrees of freedom parameters, respectively,
and we write $\varepsilon_t\sim\mathcal{ST}(0,\omega^2,\alpha,\nu)$.
When $\nu>30$ the density is effectively that of a skew-normal distribution,
whereas when $\alpha=0$ it is that of a $t$ distribution.
We fix values of $\nu$, and use the 
prior at~\eqref{eq:prior1} for $\betavec$, so that the model parameters are $\{\betavec,\tau^2,\omega^2,\alpha\}$.

Bayesian estimation of the model parameters
uses a
conditionally Gaussian representation 
of the skew-$t$. 
This introduces latent variables $\zeta_t,w_t$, 
and expresses $\varepsilon_t$ conditional on these values as 		$\varepsilon_t=\psi\zeta_t+\epsilon_t$, where 
\begin{equation*}
		\epsilon_{t}|w_t\sim\mathcal{N}(0,\sigma^2/w_{t})\,,\,
		w_{t}\sim\mathcal{G}\left(\frac{\nu}{2},\frac{\nu}{2}\right)\,,\,
		\zeta_t|w_{t}\sim\mathcal{TN}_{[0,\infty)}\left(0,1/w_{t}\right)\,,
\end{equation*}
$\mathcal{G}$ denotes a Gamma distribution, 
$\mathcal{TN}_{[0,\infty)}$ is a Gaussian truncated to $[0,\infty)$,
and $\{\alpha,\omega\}$ are re-parameterized as $\{\psi,\sigma\}$ 
with 
$\psi=\alpha\sigma$ and $\sigma^2=\omega^2-\psi^2$. Integrating out $\zeta_t,w_t$
recovers
the skew-$t$ distribution $\varepsilon_t\sim\mathcal{ST}(0,\omega^2,\alpha,\nu)$. 
Following~\cite{FruPyn2010}, the hyperpriors are $\psi\sim\mathcal{N}(0,D_0)$,
$\sigma^2\sim\mathcal{IG}(c_0,C_0)$ and $\tau^2\sim\mathcal{IG}(b_0,B_0)$, with 
constants $D_0=1$, $c_0=2.5$, $b_0=1$, $B_0=0.005$, $C_0=0.5s_y^2$ and $s_{y}^2$ 
denoting the sample variance of the response $Y_t$. 

Given $\xivec$, the
re-parameterized model parameters $\varthetavec=\{\betavec,\tau^2,\psi,\sigma^2\}$  are estimated using their Bayesian
posterior computed by an MCMC sampler that generates 
$\zetavec=(\zeta_1,\ldots,\zeta_T)'$ and $\bm{w}=(w_1,\ldots,w_T)'$, similar to those 
proposed by~\cite{panagiotelis2008} and~\cite{FruPyn2010}. As with the Gaussian
probabilistic ESN, prediction is
based on Monte Carlo draws from the ensemble density at~\eqref{eq:ensemble} 
produced using a minor adjustment of Algorithm~2. Details are given
in Part~B of the WA.
\section{\mbox{Deep Distributional Time Series}}\label{sec:copulamod}
While the deep time series models in Sec.~\ref{sec:DNN} provide probabilistic forecasts they have two drawbacks: (i)~the feature
vector only affects the mean of the response equation at~\eqref{eq:ESN}, and
(ii)~the density forecasts are not calibrated in any  manner.  
In this section, a copula model is outlined
that is a deep distributional time series model in which the 
feature vector affects the entire predictive distribution 
and the probabilistic forecasts are marginally calibrated. 
 
\subsection{Marginal calibration}
\cite{gneiting+br07} discuss different
forms of calibration of 
density forecasts, including predictive marginal calibration, 
which is defined as follows. For $t=T+1,\ldots,\infty$, assume a future observation of  $Y_t$ has true
distribution $H_{t|T}(y)$ and forecast distribution $F_{t|T}(y)$, where the subscript  
indicates
that both distributions are conditional on the filtration at time $T$. 
 Then, if 
$\bar H(y)\equiv\lim_{h_1\rightarrow \infty}\left\{\frac{1}{h_1}\sum_{h=1}^{h_1}H_{T+h|T}(y)\right\}$ and
$\bar F(y)\equiv\lim_{h_1\rightarrow \infty}\left\{\frac{1}{h_1}\sum_{h=1}^{h_1}F_{T+h|T}(y)\right\}$, the forecast
distributions are called marginally calibrated if and only if $\bar F(y)=\bar H(y)$. 

\cite{gneiting+br07}
highlight that because $H_{t|T}$ is unknown for $t>T$, marginal calibration can 
be assessed in practice by
comparing the empirical distribution function over the forecast horizon to the average of the corresponding distributional forecasts. In a study with a moving window (such as ours) and
forecast origins $T=T_{\mbox{\tiny start}},\ldots,T_{\mbox{\tiny finish}}$, there are a total of $N=T_{\mbox{\tiny finish}}-T_{\mbox{\tiny start}}+1$ sets of forecasts. Averaging over these 
and the forecast horizon of length $h_1$, 
gives
\begin{equation}
\hat{H}(y)=\frac{1}{Nh_1}\sum_{T=T_{\mbox{\tiny start}}}^{T_{\mbox{\tiny finish}}}\sum_{t=T+1}^{T+h_1}
 \mathcal{I}(y_{t}\leq y)\,,\mbox{ and }
\hat{F}(y)=\frac{1}{Nh_1}\sum_{T=T_{\mbox{\tiny start}}}^{T_{\mbox{\tiny finish}}}\sum_{t=T+1}^{T+h_1}
F_{t|T}(y)\,,
\end{equation}
where $\mathcal{I}(X)$ is an indicator function equal to one if $X$ is true, and zero otherwise.
The closer $\hat F$ is to $\hat H$,
the greater evidence for predictive marginal calibration. 
%
%
%

\subsection{Copula model}\label{subsec:copmod}

We
adopt a copula model for the joint distribution
of $\bm{Y}_{(t)}=(Y_1,\ldots,Y_t)'$, conditional on $X_{(t)}=[\xvec_1|\cdots|\xvec_t]'$ and weight configuration $\xivec$, with the density decomposition
\begin{equation}\label{eq:copmod}
p(\yvec_{(t)}|X_{(t)},\xivec)=
c_{\mbox{\tiny ESN}}(F_Y(y_1),\ldots,F_Y(y_{t})|X_{(t)},\xivec)\prod_{s=1}^{t} p_Y(y_s)\,,\; \mbox{ for } t\geq 2\,,
\end{equation}
where $\yvec_{(t)}=(y_1,\ldots,y_t)'$ and  $\uvec_{(t)}=(u_1,\dots,u_{t})'$.
The deep copula process 
has $t$-dimensional density $c_{\mbox{\tiny ESN}}(\uvec_{(t)}|X_{(t)},\xivec)$
specified below
in Sec.~\ref{sec:ESNcop}. 
The density $p_Y$ and
corresponding distribution function $F_Y$ in~\eqref{eq:copmod} are time invariant, and are estimated non-parametrically from the training data.   
This ensures in-sample marginal calibration, which will extend
to predictive calibration whenever the series is stationary. 

We stress that even though in~\eqref{eq:copmod}
the distribution $Y_t|\xvec_t$ is assumed marginally invariant with respect to $\xvec_t$, the distribution $\bm{Y}_{(t)}|X_{(t)}$
is still related to the matrix of feature vector values $X_{(t)}$ though the joint distribution.
A consequence is that for $t>T$ the
predictive density of 
$Y_t$ can be heavily dependent on the feature vector $\xvec_t$, as outlined in Sec.~\ref{subsec:preddens}.
	


\subsection{Specification of the deep copula process}\label{sec:ESNcop}

At~\eqref{eq:copmod} we employ an (implicit) copula process with density  $c_{\mbox{\tiny ESN}}$ 
constructed from the joint distribution of a  
second stochastic process $\{\tilde Z_s\}$ that follows the Gaussian probabilistic ESN
at~\eqref{eq:ESN} with $\betavec$ integrated out under the
prior~\eqref{eq:prior1}. We call $\tilde Z_s$ a ``pseudo-response'' because it is 
not observed directly, but is introduced only for specification of its
implicit copula. The $t$ observations $\tilde{\bm{Z}}_{(t)}=(\tilde Z_1,\ldots,\tilde Z_t)'$, are conditionally distributed 
\begin{equation}\label{eq:conddist}
\tilde{\bm{Z}}_{(t)}| X_{(t)},\sigma^2,\tau^2,\xivec \sim \mathcal{N}\left(\tilde{\zvec}_{(t)}; \bm{0},\sigma^2\left( I+\frac{1}{\tau^2}B_\xivec(X_{(t)})B_\xivec(X_{(t)})' \right) \right)\,,
\end{equation}
with $\tilde{\zvec}_{(t)}=(\tilde z_1,\ldots,\tilde z_t)'$ and the matrix $B_{\xi}(X_{(t)})$ is specified
at~\eqref{eq:ESNmat}, but without an intercept term (i.e. the first column of ones)
because level is unidentified in a copula.

The implicit copula of a Gaussian distribution
is called a Gaussian copula, and is constructed for~\eqref{eq:conddist}
by standardizing the distribution.
Let $\bm{Z}_{(t)}=(Z_1,\ldots,Z_t)'=\sigma^{-1}S\tilde{\bm{Z}}_{(t)}$, where 
$S=\mbox{diag}(\psi_1,\ldots,\psi_t)$ is a diagonal scaling matrix
with elements $\psi_s=(1+ \bvec_s' \bvec_s/\tau^2)^{-1/2}$, and $\bvec_s$ is the $s$-th row of $B_{\xi}(X_{(t)})$. Then $\bm{Z}_{(t)}|X_{(t)},\sigma^2,\tau^2,\xivec\sim \mathcal{N}(\bm{0},R)$ where
 \begin{equation}
R = S\left( I+\frac{1}{\tau^2}B_\xi(X_{(t)}) B_\xi(X_{(t)})' \right)
S\,,
\nonumber
\end{equation}
has ones on the leading diagonal, and is a function of $(X_{(t)},\xivec,\tau^2)$. 
Then the copula  has 
density
$c_{\mbox{\tiny ESN}}(\uvec_{(t)}|X_{(t)},\xivec,\tau^2) = 
 \frac{\phi(\zvec_{(t)};\bm{0},R)}{\prod_{s=1}^t \phi_1(z_s)}\,,\label{eq:copdens}$
 where
 $z_t=\Phi_1^{-1}(u_t)$, $\zvec_{(t)}=(z_1,\ldots,z_t)'$, $\phi_1$ is the standard normal density, and $\phi(\cdot\,;\bm{0},R)$ is the density 
 of a $\mathcal{N}(\bm{0},R)$ distribution. Notice that $\sigma^2$ does 
 not feature in $c_{\mbox{\tiny ESN}}$, which is because implicit copulas are scale free.

Because 
$c_{\mbox{\tiny ESN}}$ is conditional
on $X_{(t)}$, this is a Gaussian copula process on the feature space~\citep{Wilson2010}. 
Because 
the copula captures the dependence structure in $\{Z_t\}$, 
$\bm{x}_t$ contains past values of this process (rather than 
$\{Y_t\}$), 
and the equivalent processes for the other four regional electricity prices.
These can be computed easily
as $Z_t=\Phi^{-1}(F_Y(Y_t))$. 
We write the copula density $c_{\mbox{\tiny ESN}}$  also as a function of the unknown parameter 
$\tau^2$ and employ it in the copula model at~\eqref{eq:copmod}.

\subsection{Estimation}\label{sec:copest}
Given a configuration $\xivec$, the only unknown copula parameter is $\tau^2$,
for which we adopt the Weibull prior of~\cite{KleKne2016} with scale parameter $b_{\tau^2}=2.5$.
Direct estimation using the likelihood at~\eqref{eq:copmod}
is difficult because evaluation of
$c_{\mbox{\tiny ESN}}(\uvec|X,\xivec,\tau^2)$
requires inversion of the matrix $R$, which is computationally infeasible
for all but small sample sizes. 
This problem is solved by
expressing
the likelihood conditional on $\betavec$ as follows. For a sample of size $T$, denote 
the observations as $\yvec\equiv \yvec_{(T)}$ and feature
matrix as $X\equiv X_{(T)}$. Then the conditional likelihood 
is obtained by
a change of variables from $\bm{y}$ to $\bm{z}=(z_1,\ldots,z_T)'$, with elements
$z_t=\Phi_1^{-1}(F_Y(y_t))$, so that
\[
p(\yvec|X,\betavec,\xivec,\tau^2)= p(\zvec|X,\betavec,\xivec,\tau^2)\prod_{t=1}^T 
\frac{p_{Y}(y_t)}{\phi_1(z_t)}\\
=\phi\left(\zvec;\bm{0},SB_\xivec(X)\betavec,
S^2\right)\prod_{t=1}^T\frac{p_{Y}(y_t)}{\phi_1(z_t)}\,,
\]
which can be evaluated in $O(T)$ operations because $S$ is diagonal. Part~C of the WA
provides an MCMC sampler to generate draws from the augmented posterior of $(\betavec,\tau^2)$,
so that $\betavec$ is integrated out in a Monte Carlo manner and direct 
computation of $R$ is avoided.

\subsection{Probabilistic forecasts}\label{subsec:preddens}
As with the probabilistic ESNs in Sec.~\ref{sec:DNN}, 
an ensemble is used to integrate
over the distribution of $U,V$,
so that
the density forecast of $Y_{T+h}$ 
is again given by~\eqref{eq:ensemble}. The ensemble components $p^k$ 
 (i.e. the Bayesian posterior predictive densities at~\eqref{eq:pdenscop}) are derived from 
the copula model where $\varthetavec=\{\betavec,\tau^2\}$. To do so,
the first term in the integrand of~\eqref{eq:pdenscop} is obtained by a
change of variables from  $y_{T+h}$ to $z_{T+h}=\Phi_1^{-1}(F_Y(y_{T+h}))$, so 
that
\begin{eqnarray}
\lefteqn{p(y_{T+h}|X_{(T+h)},\varthetavec) 
= p(z_{T+h}|X_{(T+h)},\betavec,\tau^2)\frac{p_Y(y_{T+h})}{\phi_1(z_{T+h})}}\nonumber\\
&= &
\frac{1}{\psi_{T+h}}\phi_1\left(
\frac{\Phi_1^{-1}(F_Y(y_{T+h}))-\mu_{T+h}}{\psi_{T+h}}
\right) \frac{p_Y(y_{T+h})}{\phi_1\left(\Phi_1^{-1}\left(F_Y(y_{T+h})\right)\right)}\,,\label{eq:pdens2}
\end{eqnarray}
where $\mu_{T+h}=\frac{1}{\psi_{T+h}}\bvec_{T+h}\betavec$, $\psi_{T+h}=(1+\frac{1}{\tau^2}\bvec_{T+h} \bvec_{T+h}')^{-1/2}$ and $\bvec_{T+h}=(\hvec_{T+h}',(\hvec_{T+h}^2)')$ is a row vector. Notice
that the entire density at~\eqref{eq:pdens2} is a nonlinear function of
the feature vector $\xvec_{T+h}$ via $\bvec_{T+h}$, so that
$Y_{T+h}$ is not marginally invariant of $\xvec_{T+h}$ in the predictive 
distribution.
As in Sec.~\ref{sec:gaussfore}, the posterior mean $E(\varthetavec|\yvec)$ is computed from
the Monte Carlo sample and plugged in for $\betavec,\tau^2$ in~\eqref{eq:pdens2}.
Because $\xvec_t$ comprises 
past values of $\{Z_t\}$ for both the focal price, and the other four regional
prices, the integrals
over $\xvec_{T+2},\ldots,\xvec_{T+h}$ in~\eqref{eq:pdenscop} 
are computed by simulating values for $Z_{T+t}$ sequentially for all five
regions. Algorithm~4 in Part~C of the WA generates
iterates from the ensemble density for the copula model.

\section{Forecasting Comparison}\label{sec:empirical}
We illustrate the accuracy of our deep model forecasts
using half-hourly price data observed during 2019 in all five regions. 
Competing models were fit using a moving window of three months training data, starting with the period
1 February--30 April. (We start at February, rather than January, because the input vector $\bm{x}_t$ defined
below includes lagged values of the series.)
This window was advanced monthly until the period 1 September--30 November, resulting
in eight fits per model.
Forecasts
were constructed for a eight month evaluation period from 1 May to 31 December; a total
of 245 days. The origin $T$ was advanced every half hour, and probabilistic forecasts
constructed for a horizon of 24 hours (i.e. $h_1=48$ half-hours).\footnote{Because there are 245 days in the 
	evaluation period, this means there was a total of
	$245 \times 48$ forecast origins. At each forecast origin,
	a total of $48 \times 5=240$ probabilistic forecast densities were constructed for each model. Thus, there were a total of 2,822,400  probabilistic forecast densities for each model.}
 When constructing
forecasts during each month, the model parameter estimates obtained from the three preceding months training data were used.
The reason each model was only refit eight times (rather than for every one of 11,760 half-hourly forecast origins) was to reduce the computational burden of the study.
A forecast horizon up to 24 hours is necessary for the bidding and 
re-bidding process in the NEM.

\subsection{Probabilistic forecasting models}\label{sec:formod}
The following four deep time series models
were applied to each of the five  price series:
\begin{itemize}
\item[] \underline{RNN}: The Gaussian probabilistic ESN outlined in
Sec.~\ref{subsec:ESN}.
\item[] \underline{RNNST}: The skew-$t$ probabilistic ESN outlined in Sec.~\ref{subsec:skewt:ESN} with 
$\nu=7$.
\item[] \underline{RNNSN}: A skew-normal probabilistic ESN, approximated using  RNNST with 
$\nu =30$.
\item[] \underline{RNNC}: The copula model outlined in 
Sec.~\ref{sec:copulamod} that uses the deep
copula constructed from a Gaussian probabilistic ESN, along with bounded
KDE margins.
\end{itemize}

For the RNN, RNNST and RNNSN, 
the feature vector contains the $n_x=271$ elements
\[
\xvec_t=(1,\underbrace{\bm{Y}_{\mbox{\tiny{ALL}},t-1}, \bm{Y}_{\mbox{\tiny{ALL}},t-2},\ldots, \bm{Y}_{\mbox{\tiny{ALL}},t-48}}_{\text{prices in previous 24 hours}},
\underbrace{\bm{Y}_{\mbox{\tiny{ALL}},t-96},\bm{Y}_{\mbox{\tiny{ALL}},t-144},
\ldots,\bm{Y}_{\mbox{\tiny{ALL}},t-336}}_{\text{prices at same hour 2 to 7 days previously}})'\,,
\]
where
$\bm{Y}_{\mbox{\tiny{ALL}},t}= (Y_{\mbox{\tiny{NSW}},t}, Y_{\mbox{\tiny{QLD}},t}, Y_{\mbox{\tiny{SA}},t}, Y_{\mbox{\tiny{TAS}},t}, Y_{\mbox{\tiny{VIC}},t})$ is the vector of all five regional prices at hour $t$ and the first entry is the intercept.
For RNNC, the $\xvec_t$ contains values of the transformed series
$Z_t=\Phi_1^{-1}(F_Y(Y_t))$ for each of the five regions
at the same lagged time points as above.
The choice of inputs is motivated by 
previous studies~\citep{panagiotelis2008,higgs2009} that identify 
strong dependence in Australian electricity prices with those in the previous 24 hours, and 
at the same time during the previous seven days. 
Price dependence is induced by serial dependence in both supply-side factors 
and in electricity consumption~\citep{smith2000}.
Complex cross-sectional serial dependence in price is also
well-documented~\citep{panagiotelis2008,higgs2009,ignatieva2016,han2020} and
is caused by inter-regional trade in electricity between the 
five regions. 

While the deep time series models above
are trained separately for each price series, the inclusion of 
lagged prices from all regions as elements of $\xvec_t$ forms 
five-dimensional multivariate deep times series models. From each model
multi-step ahead forecast densities for
the five regional prices 
are constructed jointly via simulation using Algorithms~2 and~5  outlined
in the WA. At each forecast origin,  $2000$ Monte Carlo
draws from~\eqref{eq:ensemble} are obtained for each half hour $h=1,\dots,48$ and region $r=1,\ldots,5$.

The density $p_Y$ was estimated for each series 
using the bounded KDE implemented in MATLAB function `\texttt{ksdensity}', with
bounds set equal to the admissible prices during the forecast period. These were
a floor price of -\$1000 ($Y_{T+h}\geq 0$), and a maximum price of \$14,500 prior to 1 July ($Y_{t+h}\leq 9.582$) and \$14,700 on or after that date ($Y_{t+h}\leq 9.596$).
These densities are
given in Fig.~B in the WA. This ensures 
that the predictive densities from the copula model
at~\eqref{eq:pdenscop} are constrained
to the same range, which is important for the accurate forecasting of
prices near the bounds.
%
 In contrast, density forecasts from
the other models are unconstrained, and can have substantial mass outside the bounds. 
Thus, we simply truncate their predictive densities 
to the admissible price region and normalize.

In addition to our four proposed
deep time series models, following ~\cite{serinaldi2011} and~\cite{gianfreda2018} we 
also employ GAMLSS
 as a benchmark. 
In this framework the response $Y_t$ 
has a specified parametric distribution, where the parameters are modeled 
as flexible functions of covariates.
We used the  R package~`\texttt{gamlss}',
but found the software unstable when the ESN output layer terms were used
as covariates, and also for many of the candidate distributions supported by the package. However,
the four parameter Johnson's SU distribution
with $\xvec_t$ as linear covariates worked well,
and we include forecasts from this model. \cite{serinaldi2011} also found the JSU distribution to work
well for electricity prices, which are a challenging series with which 
to calibrate models
due to the frequent price 
spikes. 
Details on our GAMLSS implementation and experiments
are given in Part~D of the WA.

\subsection{Measuring forecast accuracy}\label{sec:metrics}
For point forecasts,
the mean absolute error (MAE)
and root mean square error (RMSE) are used to measure accuracy. 
For a univariate probabilistic  
forecast distribution $F$ there are a growing number of measures of accuracy,
many of which are listed
by~\cite{nowotarski2018}, and we compute the following four.
The first is the quantile
score function $\mbox{QS}_\alpha(F^{-1}(\alpha),y)=2({\cal I}(y<F^{-1}(\alpha))-\alpha)(\alpha-y)$, where ${\cal I}(A)=1$ if $A$ is true, 
and zero otherwise.
This is also called the ``pinball loss''~\citep{raiffa61,gneiting2011}.
While the entire quantile score function is of interest, we also consider
its value at $\alpha=0.05$ and $\alpha=0.95$ as measures of lower and upper
tail accuracy, respectively. 
The second measure is 
the continuous ranked probability
score (CRPS) \citep{gneiting+br07} defined as $\mbox{CRPS}(F,y)
=\int_0^1 \mbox{QS}_\alpha(F^{-1}(\alpha),y)d\alpha$ for observed value $y$, where
the integral is computed numerically. It is a proper scoring rule
which measures overall probabilistic forecast accuracy. 

The third measure is a loss function proposed
by~\cite{fissler2016} that is based on 
the ``value-at-risk'' and ``expected shortfall'', which
are  popular measures of financial tail risk
for low
values (i.e. financial losses). To measure
tail risk associated with
high values (which is the primary source of price risk in electricity markets) 
we employ the restatement 
of the loss function for the upper tail given by~\cite{nolde2017}. For $Y\sim F$ and $\alpha$ close 
to 1, let 
$\mbox{EL}(\alpha)\equiv
E(Y|Y\geq F^{-1}(\alpha))$ be the ``expected longrise'' (analogous to
the expected shortfall measure of lower tail risk). 
Then from~\cite[Prop.~3]{nolde2017},
the loss function is 
\begin{eqnarray}
\lefteqn{S_\alpha(F,\mbox{EL},y)= {\cal I}(y\geq F^{-1}(\alpha))\left(-G_1(F^{-1}(\alpha))+G_1(y)-G_2(\mbox{EL}(\alpha))(F^{-1}(\alpha)-y)\right)} \nonumber \\
& &+(1-\alpha)\left(G_1(F^{-1}(\alpha))-G_2(\mbox{EL}(\alpha))(\mbox{EL}(\alpha)-F^{-1}(\alpha))+G_3(\mbox{EL}(\alpha))\right)\,,\label{eq:jointtail}
\end{eqnarray}
where $G_1(x)=x$ and $G_2(x)=G_3(x)=\exp(x)$.
We compute this score for $\alpha=0.975$, and lower values indicate more
accurate upper tail forecasts of the  price distribution.
The fourth and last measure is the empirical coverage of the predictive distributions
at the 95\% level.

For each of the first three metrics, we report the weighted average of their values over the evaluation period and 
over regions, with weights equal to the proportion of total
electricity consumption in each region.\footnote{These weights are 0.3687 (NSW), 0.2355 (VIC), 0.2818 (QLD),
	0.0624 (SA) and 0.0516 (TAS).}
We call these ``system-weighted'' average metric values, and they reflect
the differing importance of 
price forecasting accuracy across regions in the NEM. For the fourth metric, the coverage
is with respect to the price distributions in all five states.

\subsection{Empirical results}
\subsubsection{Predictive distributions}
\begin{sidewaysfigure}[ht]
	\caption{Out-of-sample predictive marginal calibration}
	\begin{center}
		\centering\includegraphics[width=0.99\textwidth,angle=0]{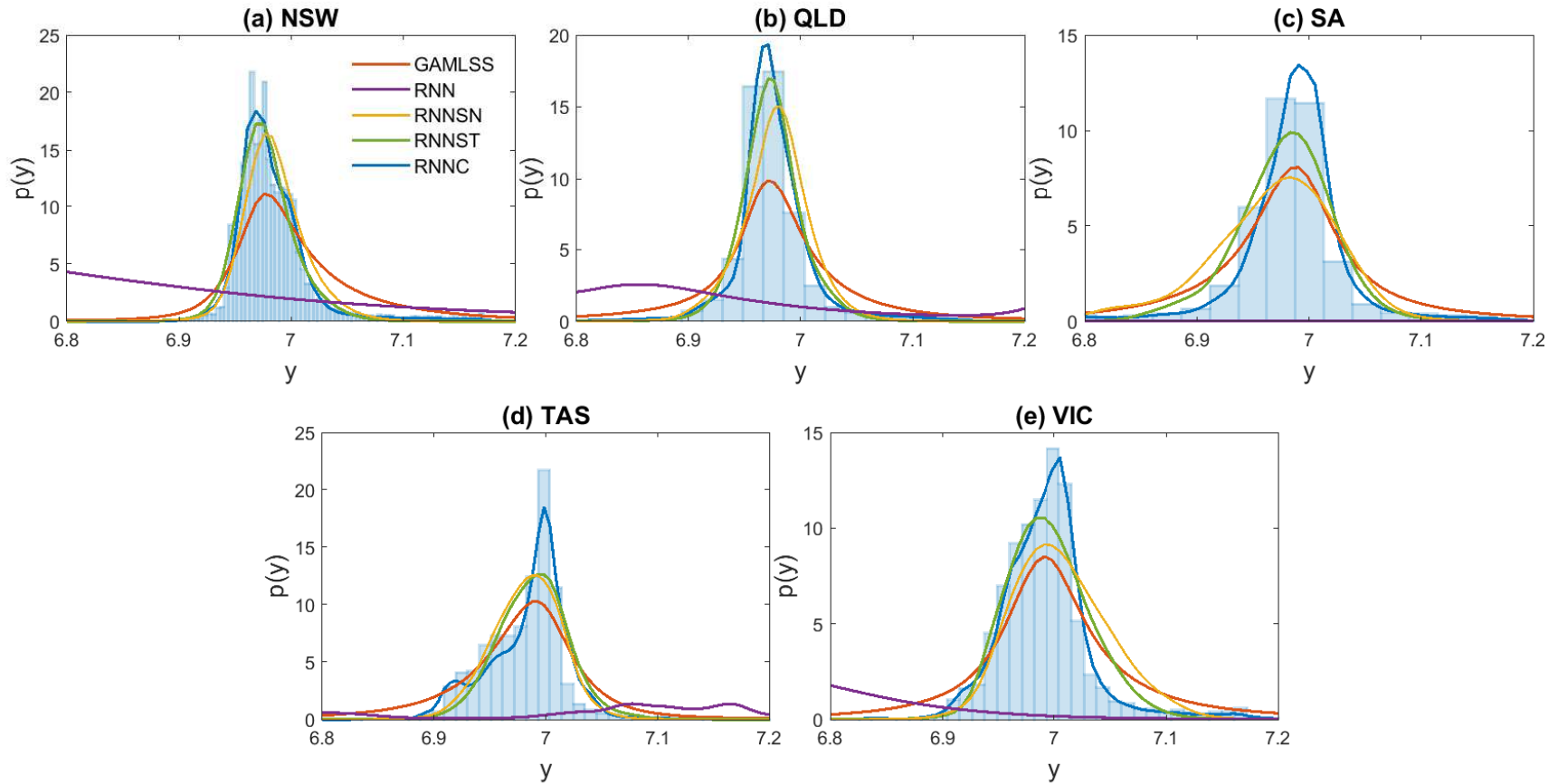}
	\end{center}
Histograms are of the out-of-sample observations
of $Y$ during
the evaluation period (1 May - 31 December 2019).
The average predictive density for 
each method is given for the four deep time series models, plus the GAMLSS 
benchmark. The deep time series copula model (RNNC) produces average 
densities close to the bounded KDEs fit to the in-sample
data found in Fig.~B of the WA by construction. The panels
correspond to forecast prices in (a) NSW, (b) QLD, (c) SA, (d) TAS and
(e) VIC. 
	\label{fig:ooscalibration}
\end{sidewaysfigure}

Fig.~\ref{fig:ooscalibration} presents histograms of the out-of-sample observations
of $Y$ for each region during
the evaluation period.  Also plotted
are the average predictive densities $\hat F$ for the five forecasting methods,
allowing a visual comparison of out-of-sample marginal calibration.
The copula model (RNNC) produces forecasts that are most accurately calibrated, followed
by RNNSN and RNNST. However, GAMLSS and RNN both exhibit poor predictive marginal calibration. 
The exceptionally wide intervals for RNN are caused by extremely inaccurate density forecasts at horizons 
close to 24 hours ahead as characterized further below.

\begin{figure}[ht]
	\caption{Prediction intervals for NSW prices at 12:00 during July 2019}
	\begin{center}
		\centering\includegraphics*[width=0.8\textwidth]{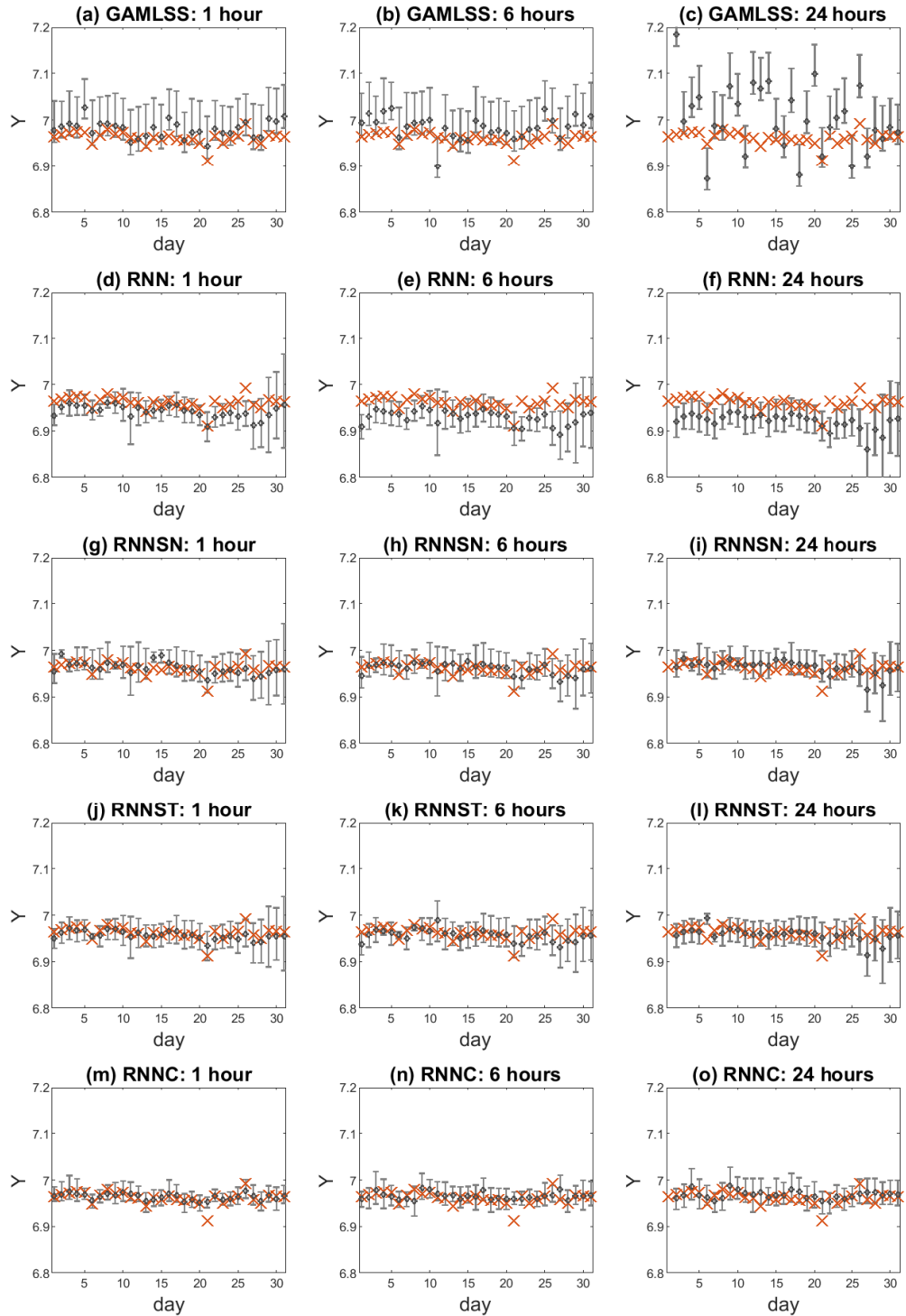}
	\end{center}
Panels correspond to forecasts from the five methods made 1, 6 and 24 hours ahead. In each panel, 95\% prediction intervals are denoted
by a vertical bar, with the predictive mean given by a grey cross. 
The true values are denoted with a red cross. Results are presented
on the logarithmic scale for clarity.
	\label{fig:NSWintervals}
\end{figure}

To visualize the predictive densities, 
Fig.~\ref{fig:NSWintervals} plots the 95\% prediction intervals
for forecasts made during the month of July 2019. Results from the five methods for forecasts 1, 6 and 24 hours ahead are given in 
separate panels. RNN forecasts are negatively biased, and for some days
lack ``sharpness''~\citep{gneiting+br07}. The latter is because a high variance is
necessary to capture the price spikes in the training sample
using a distribution with symmetric thin tails for the errors at~\eqref{eq:ESN}.
In contrast,
skew-normal disturbances (RNNSN) can produce a heavier
upper tail through right skew, but at the expense of
overly sharp densities. De-coupling of the level of asymmetry
from the kurtosis through skew-$t$ disturbances (RNNST) allows
for asymmetric predictive densities with heavy tails.
RNN, RNNSN and RNNST all have probabilistic forecasts that are nearly 
homoscedastic, because they are ensembles of
homoscedastic Bayesian predictive densities. In contrast RNNC allows for a much heavier
upper tail for electricity
prices and heteroscedasticity; the latter is because it is a distributional model.
Last, the performance of the benchmark GAMLSS model (which is also a distributional 
model) declines substantially over the forecast horizon.

\subsubsection{Forecast accuracy}
To assess forecast accuracy we compute the metrics
outlined in Sec.~\ref{sec:metrics} for each of the five methods. 
Table~\ref{tab:pointfore} reports out-of-sample point forecast accuracy, and
the asymmetric deep time series models RNNSN, RNNST and RNNC are more accurate
than RNN and GAMLSS in terms of MAE at all horizons, but not by the RMSE metric at short horizons.
At horizons of 12 or more hours, RNNC produces the most accurate point forecasts. 
%
However, the main objective is accurate probabilistic
forecasting, and Table~\ref{tab:densityfore} reports out-of-sample metrics
for this.
RNN is the poorest method by all measures, which is unsurprising given the 
asymmetry and heavy tails in even logarithmic prices.
GAMLSS is also very poor overall. 
RNNST is more accurate than RNNSN, demonstrating that allowing for heavy tails
increases accuracy over-and-above allowing for asymmetry.
However, it is the deep copula model RNNC that provides 
clearly superior forecasts at all points in the forecast horizon, both in terms
of the entire density (as measured using CRPS)
and the upper and lower tails (as measured by the quantile scores and upper tail loss function). 

\begin{figure}[ht]
	\caption{Tail risk for NSW prices at 12:00 during July 2019}
	\begin{center}
		\centering\includegraphics*[width=1\textwidth,angle=0]{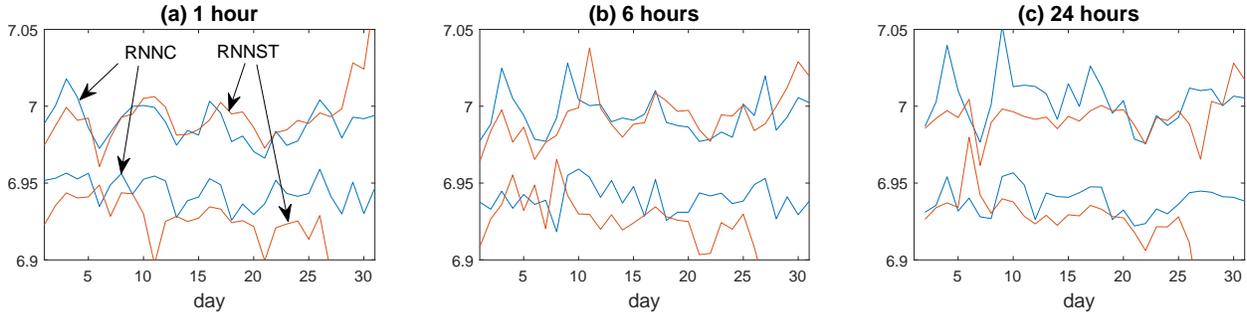}
	\end{center}
Panels (a)-(c) correspond to forecasts made 1, 6 and 24 hours ahead for the RNNC (in blue) and RNNST (in red). In each panel, both the the expected longrise
$E(Y|Y>Q(\alpha))$ at $\alpha=0.975$ and the expected shortfall $E(Y|Y<Q(\alpha))$ at $\alpha=0.025$ are plotted. Results are presented on the logarithmic scale for clarity.
	\label{fig:condE:NSW:both}
\end{figure}

Fig.~F in the WA~plots the full quantile score 
functions (pinball loss). The poor calibration of the RNN and GAMLSS probabilistic
forecasts is clear, as is the superiority of the asymmetric deep time series models (in particular RNNC).
 To compare the probabilistic forecasts for the 
two most accurate models (RNNC and RNNST) Fig.~\ref{fig:condE:NSW:both} plots
the expected shortfall ($E(Y|Y<F^{-1}(\alpha))$ for $\alpha=0.025$) and expected
longrise ($E(Y|Y>F^{-1}(\alpha))$ for $\alpha=0.975$) at 12:00 during July 2019 for
both models. The RNNC produces predictive forecasts that exhibit substantial heteroscedasticity
and are sharper (i.e. lower spread between the longrise and shortfall)
than the RNNST. These differences in the two deep time series
models are because RNNC is a distributional time series model, whereas RNNST is not.
Finally, Table~\ref{tab:RNN:covs} reports the coverage of the 
predictive distributions at the 95\% level, and there is a 
substantial difference between the methods, with the predictive distributions from RNNC exhibiting
superior coverage at all horizons.

\section{Incorporating Probabilistic Forecasts of Demand}\label{sec:demand}
We now show how to extend the deep time series models to include demand forecasts as 
additional information in a flexible fashion, allowing for greater price forecast 
accuracy.
\subsection{Role of demand forecasts}
Demand for electricity is almost perfectly inelastic to price in the short term because individual 
users face fixed tariffs. As demand varies over time, the spot price traces 
out the average supply
curve.
This explains
the occurrence of price ``spikes''
because electricity supply curves are 
typically kinked, and for short periods of time demand can exceed the location of
this kink; see~\cite{geman2006,clements2015,smith2018} and~\cite{ziel2016}.
An implication is that including accurate demand forecasts in
a price forecasting model may increase its accuracy, particularly for the upper tail. 

However, incorporating demand forecasts
into most existing time series models for price is difficult for two reasons. First,
multiple aspects of
probabilistic demand forecasts beyond the point forecast
are likely to increase the accuracy of probabilistic price forecasts. Second, demand
forecasts are likely to be related to future prices in a highly nonlinear
fashion, particularly when combined with past price values. However, both difficulties can be addressed by incorporating multiple
summaries of the demand forecast distribution as additional elements of the feature 
vector of a deep time series model.

\subsection{NSW demand forecasts}
Methods for accurate short term probabilistic forecasting of demand
are well established, with examples
for NSW demand provided by~\cite{smith2000} and~\cite{cottet2003}.
AEMO makes demand forecasts publicly available, which are
updated every half hour by an automated system. The forecasts are of
the 10th, 50th (i.e. median) and 90th percentiles of demand at a half-hourly resolution
over a horizon of one week. We employ demand forecasts for NSW 
that are exactly 
24 hours ahead,
to forecast NSW prices 24 hours ahead. Thus, only demand forecasts truly
available at the forecast origin $T$ were used.
 We label the three percentile forecasts of NSW demand as $\mbox{D}10,\mbox{D}50$ and $\mbox{D}90$. Fig.~D in the Web App.~plots actual demand against $\mbox{D}50$
 during 2019, showing the high degree of point forecast accuracy. However, empirical coverage during 2019 suggests some minor miscalibration, with
 12.1\%, 62.5\% and 95.6\% of demand observations falling below $\mbox{D}10,\mbox{D}50$ and $\mbox{D}90$, respectively. 

\subsection{NSW price forecasting study}
The forecasting study was extended, where
the NSW demand forecasts were included in the feature vectors of our two best performing deep time series models, RNNST and RNNC, for forecasting NSW price. For RNNST
\[
\xvec_t=(1,Y_{\mbox{\tiny{NSW}},t-1}, Y_{\mbox{\tiny{NSW}},t-2},\ldots,Y_{\mbox{\tiny{NSW}},t-48},
Y_{\mbox{\tiny{NSW}},t-96},Y_{\mbox{\tiny{NSW}},t-144},
\ldots,Y_{\mbox{\tiny{NSW}},t-336},
\mbox{D}10_t,\mbox{D}50_t,\mbox{D}90_t)'\,,
\]
so that $n_x=58$. The feature
vector for RNNC contained the equivalent lagged values of the transformed series $Z_{\mbox{\tiny{NSW}},t}=\Phi_1^{-1}(F_Y(Y_{\mbox{\tiny{NSW}},t}))$, and the three demand forecasts.
For comparison,
the two models without the demand forecasts (so $n_x=55$) are also used.

\begin{figure}[ht]
	\caption{Density forecast of NSW price with and without demand forecast information}
	\begin{center}
		\includegraphics*[width=0.7\textwidth,angle=0]{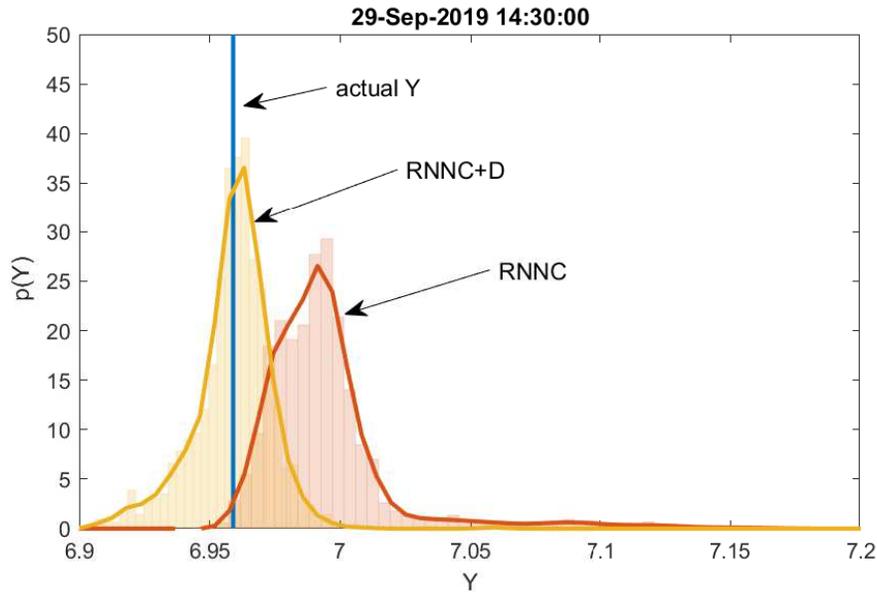}
	\end{center}
Density forecast of NSW electricity price for 29-Sep-2019 at 14:30, where the inclusion of demand forecast information greatly improved
accuracy (as measured by CRPS). The density forecast without demand forecast information
is labelled as ``RNNC'', and with as ``RNNC+D''. The actual price is marked with a 
blue vertical line. 
	\label{fig:NSWdensities2}
\end{figure}

Table~\ref{tab:foredemand} reports the average value of the forecast metrics
over the evaluation period. 
Including the demand forecasts as features decreases the accuracy of the 
point forecasts and the CRPS metric for both models. However, 
upper tail forecast accuracy increased significantly for the RNNC. This is
important in this application because
short periods of very high prices often dominate the profitability of participants 
in the NEM.
To illustrate,
Figure~\ref{fig:NSWdensities2} plots the density forecast where
the inclusion of the demand forecast information greatly increased accuracy
(other examples are given in 
Fig.~G of the WA). 

Finally, we also experimented with employing the ESN at~\eqref{eq:ESN} without the quadratic term
in the output layer, but found that this did decrease forecast accuracy significantly. 
Moreover, including ridge priors with different values of $\tau^2$ for the linear
and quadratic terms in the output layer had little effect on forecast accuracy; see Part~F of the WA for further details.


\section{Conclusion}\label{sec:conc}
This paper makes contributions to both the time series and energy economics
literatures.
We propose new deep time
series models that exploit the reservoir computing techniques
found in ESNs for high frequency time series.
Recasting a quadratic ESN as a statistical time series model, with 
a shrinkage prior for regularization of the output layer weights, allows for uncertainty 
quantification and 
probabilistic forecasting. 
Our
first approach is to allow for different error distributions for the output layer,
extending the
Bayesian methodology of~\cite{mcdermott+w17,mcdermott2019deep}.
However, our main methodological contribution is the proposal of a new
deep {\em distributional} time series model. This is obtained 
by constructing the implicit copula of a Gaussian probabilistic ESN,
and extends the deep distributional
regression method of~\cite{KleNotSmi2019} to time series.
This copula is a deep process on the feature space, allowing for 
highly adaptive nonlinear serial dependence, and generalizing existing echo state Gaussian processes~\citep{chatzis2011echo}. When combined with a nonparametric estimate
of $p_Y$, it allows for marginal calibration. The entire density forecast is a 
function of the feature vector through~\eqref{eq:pdens2}, unlike with
the other deep time series models. Our empirical work also suggests that accurate in-sample
marginal calibration also results in more accurate calibration of both the out-of-sample 
marginal and predictive distributions.

While the distribution of intraday electricity prices have 
a sizable predictable component, these time series
are complex~\citep{ignatieva2016,manner2019}. 
Key to their accurate modeling and forecasting is to capture 
jointly three features: nonlinear serial
dependence, high levels of 
asymmetry and kurtosis, and strong time-variation in the distribution.
All our proposed deep time series models account for the first feature, while
allowing for skew-normal or skew-$t$ disturbances also accounts for the second.
However, only the deep copula model allows for all three, 
producing the most accurate probabilistic forecasts.
 Recent models~\citep{ziel2016,shah2020} 
suggest that including demand forecasts may further increase probabilistic
forecasting accuracy. We find this to be the case
for the upper tail of the 24 hour ahead NSW electricity prices
when including demand forecasts as features in the deep copula model.

Last, we highlight two areas for future work. First, estimation
of our deep time series models using variational methods, rather than 
MCMC, has the potential to speed up computations.
Second, there are other
applications of our
deep time series models. One example is macroeconomic forecasting, 
where large regularized 
time-varying parameter models are popular~\citep{bitto2019,carriero2019,huber2020}.
The deep time series models provide a flexible alternative,
as they can incorporate  
 feature vectors of many economic variables regularized through
the reservoir structure of the ESN, plus Bayesian shrinkage of the output layer
coefficients. Moreover,
the copula model can produce time-varying and asymmetric probabilistic forecasts, which
is a feature of this problem.
\newpage
\singlespacing
\bibliography{references}
\begin{table}[htb]
	\begin{center}
	\small
	\caption{Point forecast accuracy of models during the evaluation period}\label{tab:pointfore}
	\begin{tabular}{llllllll}
		\hline
		&\multicolumn{7}{c}{Hours Ahead in the Forecast Horizon}\\ \cline{2-8}
		Model & 0.5 hour & 1 hour & 2 hours & 3 hours & 6 hours & 12 hours & 24 hours \\ 
		\hline
		&\multicolumn{7}{l}{MAE}\\ \cline{2-8}
GAMLSS & 0.0226$^{***}$ & 0.0314$^{***}$ & 0.0408$^{***}$ & 0.0468$^{***}$ & 0.0591$^{***}$ & 0.0692$^{***}$ & 0.0766$^{***}$ \\ 
  RNN & 0.0249$^{***}$ & 0.0319$^{***}$ & 0.0411$^{***}$ & 0.0494$^{***}$ & 0.0600$^{***}$ & 0.0722$^{***}$ & 0.0800$^{***}$ \\ 
  RNNSN & 0.0221$^{***}$ & 0.0219$^{*}$ & 0.0239 & 0.0252 & 0.0262 & 0.0268$^{**}$ & 0.0278$^{***}$ \\ 
  RNNST & 0.0189 & 0.0210 & 0.0231 & 0.0242 & 0.0250 & 0.0255$^{***}$ & 0.0257$^{***}$ \\ 
 \emph{RNNC}& \emph{0.0169} & \emph{0.0194} & \emph{0.0224} & \emph{0.0238} & \emph{0.0252} & \emph{0.0244} & \emph{0.0239} \\ 
 
		&\multicolumn{7}{l}{RMSE}\\ \cline{2-8}
GAMLSS & 0.0340$^{\mbox{\tiny{+}}}$ & 0.0461$^{\mbox{\tiny{++}}}$ & 0.0610$^{\mbox{\tiny{+++}}}$ & 0.0623$^{\mbox{\tiny{+++}}}$ & 0.0644$^{\mbox{\tiny{+++}}}$ & 0.0653 & 0.0661 \\ 
  RNN & 0.0537 & 0.0616$^{\mbox{\tiny{**}}}$ & 0.0712 & 0.0797 & 0.0935 & 3.7173$^{***}$  & 16.5496$^{***}$  \\ 
  RNNSN & 0.0538 & 0.0572 & 0.0603 & 0.0615 & 0.0623 & 0.0629 & 0.0638 \\ 
  RNNST & 0.0528$^{\mbox{\tiny{+}}}$ & 0.0571 & 0.0603 & 0.0612 & 0.0623 & 0.0630 & 0.0635 \\ 
  \emph{RNNC} & \emph{0.0538 }& \emph{0.0573 }&\emph{ 0.0613} & \emph{0.0638} & \emph{0.0657} & \emph{0.0632} & \emph{0.0629} \\ 
  
		\hline
	\end{tabular}
\end{center}
		Both the system-weighted mean absolute error (MAE) and root
		mean squared error (RMSE) are reported. Results are given for the four deep time series models, plus GAMLSS. Results of a two-sided Diebold-Mariano test of the equality of the mean error of the RNNC with each of the other models are reported. Rejection of the null hypothesis of equal means at the 10\%, 5\% and 1\% levels are denoted by ``*'', ``**'' and ``***'' if favorable to the RNNC model, or by ``${\mbox{\tiny{+}}}$'', ``${\mbox{\tiny{++}}}$'' and
		``${\mbox{\tiny{+++}}}$'' if unfavorable.
\end{table}

\begin{table}[ht]
\begin{center}
\caption{Probabilistic forecast accuracy during the evaluation period.}\label{tab:densityfore}
\centering\small
\begin{tabular}{lllllllll}
  \hline
		&\multicolumn{7}{c}{Hours Ahead in the Forecast Horizon}\\ \cline{2-8}
		Model & 0.5 hour & 1 hour & 2 hours & 3 hours & 6 hours & 12 hours & 24 hours  \\ \hline
  &\multicolumn{7}{l}{{\em CRPS}}\\ \cline{2-8}
GAMLSS & 0.0180$^{***}$ & 0.0252$^{***}$ & 0.0335$^{***}$ & 0.0392$^{***}$ & 0.0508$^{***}$ & 0.0601$^{***}$ & 0.0652$^{***}$ \\ 
  RNN & 0.0203$^{***}$ & 0.0235$^{***}$ & 0.0299$^{***}$ & 0.0364$^{***}$ & 0.0449$^{***}$ & 0.0574$^{***}$ & 0.4061$^{***}$ \\ 
  RNNSN & 0.0174$^{***}$ & 0.0177$^{**}$ & 0.0191 & 0.0201 & 0.0207 & 0.0212$^{**}$ & 0.0219$^{***}$ \\ 
  RNNST & 0.0157$^{***}$ & 0.0172$^{***}$ & 0.0187$^{***}$ & 0.0195 & 0.0201 & 0.0206$^{***}$ & 0.0208$^{***}$ \\
 \emph{RNNC} & \emph{0.0135} & \emph{0.0154} & \emph{0.0177} & \emph{0.0189} & \emph{0.0199} & \emph{0.0193} & \emph{0.0187} \\ 
  &\multicolumn{6}{l}{{\em Joint Upper Tail Loss at }$\alpha=0.975$}\\ \cline{2-8}
GAMLSS & 31.6394$^{*}$ & 32.4717$^{***}$ & 33.2367$^{***}$ & 33.9396$^{***}$ & 35.9352$^{***}$ & 40.0541$^{***}$ & 39.7960$^{***}$ \\ 
  RNN & 30.8568 & 31.5025 & 33.3171 & 35.5605 & 39.2732$^{**}$ & 41.9374$^{***}$ & 33.4037$^{***}$ \\ 
  RNNSN & 32.1052$^{**}$ & 31.2797 & 31.6342 & 31.6492 & 31.7099 & 31.5852$^{***}$ & 31.3780$^{***}$ \\ 
  RNNST & 31.3493$^{*}$ & 31.4414 & 31.6911 & 32.0056$^{*}$ & 31.8607 & 32.2372$^{***}$ & 31.8694$^{***}$ \\ 
  \emph{RNNC} & \emph{30.1002} & \emph{30.6244} & \emph{31.0090} & \emph{31.0934} & \emph{31.1121} & \emph{30.6238} & \emph{29.7661} \\ 
  
&\multicolumn{7}{l}{{\em Quantile Score at} $\alpha=0.95$}\\ \cline{2-8}
GAMLSS & 0.0056$^{***}$ & 0.0069$^{***}$ & 0.0082$^{***}$ & 0.0093$^{***}$ & 0.0124$^{***}$ & 0.0190$^{***}$ & 0.0198$^{***}$ \\ 
  RNN & 0.0050 & 0.0057 & 0.0078$^{***}$ & 0.0107$^{***}$ & 0.0155$^{***}$ & 0.0201$^{***}$ & 1.0808$^{***}$ \\ 
  RNNSN & 0.0056 & 0.0050 & 0.0054 & 0.0055 & 0.0057 & 0.0058$^{*}$ & 0.0060$^{***}$ \\ 
  RNNST & 0.0048 & 0.0050 & 0.0054 & 0.0057$^{**}$ & 0.0057$^{*}$ & 0.0062$^{***}$ & 0.0063$^{***}$ \\ 
  \emph{RNNC} & \emph{0.0036} & \emph{0.0043} & \emph{0.0048} & \emph{0.0050} & \emph{0.0052} & \emph{0.0050} & \emph{0.0046} \\ 
  
 &\multicolumn{7}{l}{{\em Quantile Score at} $\alpha=0.05$}\\ \cline{2-8}
 
GAMLSS & 0.0049$^{***}$ & 0.0080$^{***}$ & 0.0129$^{***}$ & 0.0167$^{***}$ & 0.0242$^{***}$ & 0.0260$^{***}$ & 0.0298$^{***}$ \\ 
  RNN & 0.0047 & 0.0053 & 0.0057$^{***}$ & 0.0061 & 0.0069$^{***}$ & 0.0096$^{***}$ & 0.5602$^{***}$ \\ 
  RNNSN & 0.0046$^{**}$ & 0.0055 & 0.0056 & 0.0062 & 0.0061 & 0.0064 & 0.0065 \\ 
  RNNST & 0.0048$^{***}$ & 0.0055$^{*}$ & 0.0059$^{**}$ & 0.0061 & 0.0064 & 0.0063 & 0.0066 \\ 
  \emph{RNNC} & \emph{0.0039} & \emph{0.0046} & \emph{0.0053} & \emph{0.0058} & \emph{0.0061} & \emph{0.0059} & \emph{0.0057} \\  
  \hline
\end{tabular}
\end{center}
System-weighted metrics are reported for the four deep time series models, plus
GAMLSS, at different points in time during a 24 hour forecast horizon. 
Lower values suggest greater accuracy. Results of a two-sided Diebold-Mariano test of the equality of the mean error of the RNNC with each of the other models are reported.
Rejection of the null hypothesis of equal means at the 10\%, 5\% and 1\% levels are denoted by ``*'', ``**'' and ``***'' if favorable to the RNNC model, or by ``${\mbox{\tiny{+}}}$'', ``${\mbox{\tiny{++}}}$'' and
``${\mbox{\tiny{+++}}}$'' if unfavorable.
\end{table}

\begin{table}[ht]
\begin{center}
\centering\small\renewcommand\arraystretch{1.25}
 \caption{Coverage of 95\% prediction intervals during the 
	evaluation period.}\label{tab:RNN:covs}
\begin{tabular}{lccccccc}
  \hline
		&\multicolumn{7}{c}{Hours Ahead in the Forecast Horizon}\\ \cline{2-8}
		Model & 0.5 hour & 1 hour & 2 hours & 3 hours & 6 hours & 12 hours & 24 hours  \\ \hline
GAMLSS & 0.9078 & 0.8148 & 0.7278 & 0.6928 & 0.6476 & 0.5838 & 0.5208 \\ 
  RNN & 0.8715 & 0.8609 & 0.7906 & 0.7173 & 0.6336 & 0.5698 & 0.6124 \\ 
  RNNSN & 0.8496 & 0.8386 & 0.8354 & 0.8187 & 0.8123 & 0.7998 & 0.7759 \\ 
  RNNST & 0.8550 & 0.8424 & 0.8353 & 0.8216 & 0.8171 & 0.7955 & 0.7667 \\ 
  \emph{RNNC} & \emph{0.9279} & \emph{0.9012} & \emph{0.8791} & \emph{0.8741} & \emph{0.8620} & \emph{0.8594} & \emph{0.8706} \\ 
   \hline
\end{tabular}
\end{center}
  Results are reported
 	for the four deep time series models, plus
 	GAMLSS, at different points in time in the 24 hour forecast horizon.
 	Values closer to 0.95 indicate improved calibration of the probabilistic forecasts.
\end{table}

\begin{table}[htb]
	\begin{center}
\centering\small\renewcommand\arraystretch{1.25}
 \caption{NSW price forecast accuracy when
 	including demand forecast information.} \label{tab:foredemand}
\begin{tabular}{lllllllll}
  \hline\hline
Model &  MAE & RMSE & CRPS & &JS &QS95 & QS05 & C95 \\ 
  \hline
&   \multicolumn{3}{l}{Point and Density Accuracy} &  &\multicolumn{3}{l}{Tail Accuracy} & \\ \cline{2-4} \cline{6-8}
\textit{RNNC+D} & \textit{0.0192 } &  \textit{0.0314}  &  \textit{0.0146} & & \textit{30.6902}  &  \textit{0.0045}   & \textit{0.0025} & 82.6\% \\
RNNC & 0.0188$^{\mbox{\tiny{+++}}}$   & 0.0311$^{\mbox{\tiny{+++}}}$ &  0.0144$^{\mbox{\tiny{+++}}}$ &&  31.0350$^{**}$  &  0.0048$^{***}$    & 0.0023 & 84.5\% \\   
 RNNST &0.0250$^{***}$    &0.0401    &0.0221$^{***}$ &  &39.4411$^{***}$    &0.0121$^{***}$    &0.0049$^{***}$ & 46.2 \%\\
  RNNST+D & 0.0256$^{***}$ &   0.0395  &  0.0222$^{***}$ && 39.6431$^{***}$   & 0.0123$^{***}$  &  0.0046$^{***}$ & 53.9\% \\ 
   \hline\hline
\end{tabular}
\end{center}
 Accuracy of 24 hour ahead NSW (logarithmic) price forecasts during the evaluation period. Models that include 24 hour ahead demand forecast information as features are labelled as ``+D''. The metrics are described in the text, and
include the joint score (JS) at $\alpha=0.975$, quantile score (QS) at $\alpha=0.95,0.05$, and the empirical coverage of the 95\% prediction intervals (C95). 
Results of a two-sided Diebold-Mariano test of the equality of the mean error of the RNNC+D with the other methods are reported. Rejection of the null hypothesis of equal means at the 10\%, 5\% and 1\% levels are denoted by ``*'', ``**'' and ``***'' if favorable to the RNNC+D model, or by ``${\mbox{\tiny{+}}}$'', ``${\mbox{\tiny{++}}}$'' and ``${\mbox{\tiny{+++}}}$'' if unfavorable.
\end{table}




\end{document}